\documentclass{JHEP3}

\input{epsf}
\usepackage{epsfig}
\usepackage{amssymb}
\usepackage{amsfonts}
\usepackage{amsbsy}

\newtheorem{conjecture}{Conjecture}

\def\AdS{{AdS}}

\newcommand{\be}{\begin{equation}}
\newcommand{\ee}{\end{equation}}
\newcommand{\ba}{\begin{array}}
\newcommand{\ea}{\end{array}}
\newcommand{\bea}{\begin{eqnarray}}
\newcommand{\eea}{\end{eqnarray}}

\def\Vol{\mbox{Vol}}
\def\norm#1{\left\Vert {#1} \right\Vert}

\def\IZ{\mathbb{Z}}

\def\half{\frac{1}{2}}

\newcommand{\eq}[1]{Eq.~(\ref{eq:#1})}

\def\tg{{\tilde g}}

\def\tnabla{{\tilde \nabla}}
\def\tDelta{{\tilde \Delta}}
\def\tR{{\tilde R}}

\def\one{{\hbox{ 1\kern-.8mm l}}}

\def\ba{\bar{a}}

\title{Effective potential and warp factor dynamics}

\author{Michael R. Douglas$^{1,\&}$\\
$^1$Simons Center for Geometry and Physics\\
Stony Brook University\\
Stony Brook, NY 11794 USA\\
\\
$^\&$I.H.E.S., Le Bois-Marie, Bures-sur-Yvette, 91440 France\\
{\tt douglas@max2.physics.sunysb.edu}
}

\abstract{
We define an effective potential describing all massless and massive
modes in the supergravity limit of string/M theory compactification which is valid off-shell,
{\it i.e.} without imposing the equations of motion.
If we neglect the warp factor, it is unbounded below, as
is the case for the action in Euclidean quantum gravity.
By study of the constraint which determines the warp factor, 
we solve this problem, obtaining
a physically satisfying and tractable description of the dynamics of the warp factor.
}

\begin{document}

\section{Introduction}

If our universe is described by string/M theory, there exist six or seven extra dimensions
of space, not yet detected by experiment.  This is possible because the extra dimensions can
take the form of a small, compact manifold $X$.   A basic question in string/M theory is
to know what types of manifold are allowed, and to
find general relations between their geometry and physics.  
In the present state of the art, this is generally done by
solving the equations of motion for the supergravity theory which describes the
low energy limit, and then taking into account various stringy and quantum effects.

The first works on string compactification assumed that some supersymmetry
is preserved at the compactification scale, for good physical and
mathematical reasons.  Well known physical arguments suggest low energy supersymmetry;
supersymmetry favors stability; and supersymmetry places strong constraints on the local
geometry of $X$.  In the best studied cases, $X$ is a complex K\"ahler manifold,
so powerful techniques of algebraic geometry are available for the analysis.

More recently, there has been a shift towards techniques which do not assume supersymmetry.   
After all, supersymmetry is broken in our universe, and we don't know at what scale it is broken.
String/M theory suggests many other solutions to the hierarchy problem,
such as large extra dimensions \cite{Arkani-Hamed:1998rs} or warped hierarchies  
\cite{Randall:1999ee}.
Or, the hierarchy might simply be a chance property, following from a lucky tuning of parameters
in a small subset of vacua.  Given a measure factor (a probability distribution over vacua), presumably
emerging from early cosmology, it might turn out that this class of compactifications outweighs
those with low energy supersymmetry 
\cite{Douglas:2004qg,Susskind:2004uv}.

Another reason to be interested in compactifications without supersymmetry is to get
models of inflation, because the required positive vacuum energy breaks supersymmetry.  Indeed
this is a reason to study not just vacua, meaning long-lived compactifications to maximally symmetric
space-times, but the dynamics on the larger configuration space which contains the vacua.  

At present the main technique for addressing any of these questions is to start from a class
of compactifications, say with a specific choice of topology for the compact manifold $X$, and
derive a four (or $d$) dimensional effective field theory which describes theories within this class.
One begins by identifying various ``pseudo-moduli fields''
such as metric moduli, brane positions and the like, which parameterize the vacua within this class.
One then derives an effective potential, which is a function of the these fields.
This is usually done by combining various exact results for
related supersymmetric compactifications, brane world-volume theories, and supersymmetric
quantum field theories, in a first approximation by adding them, and then considering
corrections. 

Having derived an effective potential, one then looks for its local minima.
The first issue is that, since the potential always goes to zero at
large volume and weak coupling \cite{Dine:1985he,Giddings:2003zw},
one must find effects that produce a barrier to this
runaway.  
Having done this, the easiest way to argue that local minima exist at finite moduli is to
show that,
to a good approximation, the effective potential in the regime under study is a sum of
many different, loosely correlated, contributions from different sources: fluxes, branes,
quantum effects, curvature, and so on.  By approximately balancing two or three of these
contributions and then tuning their precise strengths,
one can obtain potentials with local minima.  

One important conclusion from this work is that there is no particular obstacle within
string theory to realizing de Sitter space-time \cite{Kachru:2003aw}, and thus the small
observed dark energy could be a cosmological constant.
Indeed, given that one can obtain local minima with a small negative
vacuum energy, compared to the individual contributions, it would seem that only an incredible
conspiracy between the different effects would prevent one from obtaining similar vacua with small
positive vacuum energy.  In relatively simple and controllable models such as
the explicit KKLT realizations of \cite{Denef:2004dm},  one could in principle get enough control to 
rule out such conspiracies.  In practice, one brings in some physical intuition at this point,
using arguments such as parametric separation of energy scales of different effects (so they
cannot always cancel) or spatial separation of effects from different regions in $X$.  Combining these
arguments with the lack of any proposed mechanism for the supposed conspiracy,  the conclusion
seems well enough justified, though better arguments would certainly be welcome.

This general approach and many examples are reviewed in \cite{Douglas:2006es}.
It is very effective in determining general structure and relations between parameters such
as masses and coupling constants; for example the general differences between IIb, IIa, large
volume and heterotic compactifications are all readily understandable in these terms.
One can also show that certain classes of models cannot contain
vacua, or cannot realize slow-roll inflation \cite{Hertzberg:2007wc},
because of runaways to large volume or weak coupling.
Other reasons to develop this approach are that it could in principle be generalized to nongeometric
compactifications, and finally it is the best we can hope to do for the vast majority of compactifications.

Once one grants the validity of the effective action, the question of the existence of vacua with
positive vacuum energy, and even general nonsupersymmetric vacua, becomes in principle
straightforward to answer.  It also leads to a simple picture of dynamics.  The best studied example is
the dynamics of inflation, which can be realized by slowly rolling down a nearly flat potential.
More recently, models have been suggested which take advantage of other structure, such
as in the kinetic terms \cite{Alishahiha:2004eh}.  Although one expects the
effective action picture to break down
at high energies, in the best case at the (higher dimensional) Planck scale, this can still be well
above the energy scale of slow-roll inflation.

One problem with the effective potential approach
is that string theory and even simple Kaluza-Klein reduction
involve an infinite set of fields, while the usual truncation to the pseudo-moduli fields is
somewhat {\it ad hoc}.  Although there is a simple argument that one can solve the equations
of motion for massive fields in terms of light fields, this ignores the possibility
of tachyonic modes, which will destabilize vacua.  A candidate vacuum for present-day physics
must be tachyon-free, while configurations with tachyons are important for cosmology.  
In dynamical situations in which fields undergo large variations, of course
the splitting of fields into ``massless'' and ``massive'' can change drastically.

There are simple arguments that massive modes are under control,
but these tend to use supersymmetry.  For example,
one often builds up an effective potential by combining sectors which individually respect
different supersymmetries.
Another simple
argument is that, even after supersymmetry breaking, the effective potential is approximately
a sum of squares, up to corrections of order $F/M_{Planck}^2$.  Such arguments seem believable
given a hierarchy between the supersymmetry breaking scale and the masses of the new modes,
but are not convincing otherwise.

A second level of analysis in which one can see the massive modes
is to solve the full $9+1$ or $10+1$-dimensional
supergravity field equations, or the beta function conditions for a conformal
world-sheet theory, or perhaps even the full string field theory equations of motion.
These various approaches include more and more modes at the cost of eventual intractability.
Of course, one does not need an exact solution for all the modes; even the analysis 
of linearized fluctuations of massive modes around a solution would
go a long way towards answering the stability question.

A problem with these approaches is that they are classical (from the point of
view of space-time), while the existing constructions of vacua usually rely on quantum corrections for
stability and supersymmetry breaking.  To try to address this problem, one can note that,
despite being formulated in higher dimensions or with string fields, we can still think of these
approaches as each based on an effective potential, essentially defined as the higher dimensional
part of the action, but now taken as a function of all of the massless
and massive modes.  The higher dimensional equations of motion correspond to solving the
condition $\partial V/\partial \phi^i=0$, while the linearized stability analysis simply 
corresponds to computing the mass matrix $\partial^2 V/\partial \phi^i\partial \phi^j$.
To the extent that one can make this idea precise, one can then try to deal with quantum
corrections by the same prescription which was used before, namely to simply add them in,
or make other corresponding adjustments of the effective potential.  We could refer to
this entire class of approaches as ``semiclassical,'' to be contrasted with a fully
nonperturbative approach such as gauge-gravity duality, which unfortunately is not known
to exist for theories with positive vacuum energy.

Since in the supergravity limit, a vacuum is a solution of well understood higher
dimensional equations
of motion, the problem of constructing an effective potential which takes massive
modes into account is in principle just one of isolating the appropriate terms in
the original action.  Following up on works such as \cite{Taylor:1999ii,Giddings:2001yu,deAlwis:2003sn},
in \cite{Giddings:2005ff} Giddings and Maharana proposed an effective 
potential, based on reinterpreting the constraint in the Einstein equations.
They went on to show that it worked in examples such as that of \cite{Giddings:2001yu}.

While we did not know about this proposal during the course of our work,
and only found out about it after having distributed our work as a 
preprint, in general we followed a similar approach, and obtained an 
effective potential consistent with that of \cite{Giddings:2005ff}, but with many differences
in our discussion.
We show explicitly that a critical point of the potential
solves all the equations of motion,  
and we have eliminated the implicit assumption
that space-time is Minkowski made at various points in \cite{Giddings:2005ff}.  
Another difference is that the two proposals
use different conventions for the warp factor.  The conventions followed
here have the great advantage that the constraint becomes {\bf linear}, enabling us to
go on and better understand the physics in \S \ref{s:physical} and \S \ref{s:stability}.

On further consideration of the resulting effective potential, one realizes that 
there is a significant problem with its interpretation, which we now explain.

\subsection{Questions addressed in this work}

In this work we develop the analysis of compactifications starting from the
equations of motion of general relativity coupled to the type of matter sectors which
appear in maximal supergravity theories, including possible stringy corrections.
We believe that the same ideas would apply to any of the semiclassical approaches, if we
understood the relevant configuration spaces.

Some issues which we try to shed light on include:
\begin{itemize}
\item The definition of the four (or $d$) dimensional effective potential.  
While it is in principle clear how to define this for a compactified field theory (not including gravity),
this is not the case once gravity is included.  Indeed, there
are well-known difficulties in making a global definition of energy in general relativity 
\cite{Wald:1984rg}.
The usual response in this context (as in \cite{Douglas:2006es}) is to work in approximately
asymptotically flat backgrounds, and appeal to the standard definition for this case.  This 
is obviously not satisfactory when doing cosmology.

One particularly interesting contribution to the effective potential
is minus the integral of the scalar curvature of the compactification
manifold,
\be \label{eq:k-R-pot}
V_{eff} = - \half\int \sqrt{g}R^{(k)} .
\ee
This term can be obtained by the general procedure we just discussed, of regarding
the higher dimensional part of the action as an effective potential,
as we will review in section \S \ref{s:general} (see \eq{eff-cc}).
At least naively, it is responsible for anticorrelating the curvature of the
compactification manifold with that of space-time, as in the well-known $\AdS_p\times S^q$
solutions.  While there are additional terms in the Einstein
equation and this is oversimplified, at first sight it seems reasonable to think of this as one of 
the many terms in the effective potential.

However, a serious problem with this interpretation of \eq{k-R-pot} is that the integral of the
scalar curvature can be made arbitrarily large and positive, by making a rapidly varying
conformal transformation of the metric.  Thus, \eq{k-R-pot} is unbounded below, and
does not even have local minima.
This problem is closely
related to the fact that the action in Euclidean gravity is unbounded below, and to the
``wrong sign kinetic term'' for the conformal factor which may be familiar from
minisuperspace treatments of gravity.
As such, one would expect that it is solved (at least classically) by imposing the appropriate
constraints.  In particular, in the canonical formulation, the Hamiltonian constraint must
be imposed at each point in space, and determines the conformal factor in terms of the
other fields.  At least in asymptotically flat space-times, the resulting energy would be
non-negative, by the positive energy theorem
\cite{Schoen:1979zz,Witten:1981mf}.

However, it is not so clear
how the Hamiltonian constraint solves this problem, because
it directly determines only the conformal factor
on $D-1$-dimensional space, leaving the conformal factor on $X$ free to vary.  
It corresponds to the ``warp factor'' constraint in the existing analyses, 
which is a clue to the physical interpretation.

Is the potential
\eq{k-R-pot} bounded below, and if so why?  If it is, what determines the correct conformal factor
with which to evaluate it?  Can we solve for this variable to simplify the potential?

\item The nature of the warp factor constraint.  In doing Kaluza-Klein reduction,
the $d$-dimensional part of the Einstein equation (in fact, the Hamiltonian constraint)
turns into a partial differential equation for the warp factor.  By analysis of this
equation, in \cite{deWit,Maldacena-Nunez}
it was shown that one cannot realize de Sitter space-time in pure supergravity, {\it i.e.}
without stringy corrections or singularities.  Later 
it was argued  \cite{Giddings:2001yu,DeWolfe:2005uu}
that this could be evaded
using orientifolds and/or ``$0$-form flux'' (the Romans mass term in IIa supergravity),
and this is no longer considered a major obstacle.  

As the warp factor plays such an important role in the physics, perhaps even solving
the hierarchy problem, 
it would be nice to have a better conceptual understanding of it.  
Are there other conditions for this constraint to have a solution?  What if the
various sources to it are large, or widely separated on $X$ ?  What if the sources
evolve in time?

\item The possibility of compactification on negatively curved manifolds, and the nature
of flux or other effects needed to stabilize negative curvature.  
According to \eq{k-R-pot}, negative scalar curvature makes a positive contribution to the
vacuum energy, which could make de Sitter space easier to realize.  This 
point has been particularly
emphasized by Silverstein, who has proposed a variety of constructions 
of this type \cite{Saltman:2004jh,Silverstein:2007ac}.  
There are many other works on the subject, including \cite{Danielsson:2009ff}.

Now, the no-go theorems of \cite{deWit,Maldacena-Nunez} do not make any assumption
about the curvature, and would hold in this case as well.  Thus, even with negative curvature,
one still needs to call on stringy effects to get de Sitter compactifications.  Is negative
curvature actually relevant for this,  and if so why?  Perhaps consistent
compactifications of this type are always dual to others of more familiar types?

The possibility of compactification on negatively curved manifolds raises another point,
which is that there are far more of these than manifolds of zero or positive curvature.
This is illustrated by the familiar example of Riemann surfaces, for which the curvature is 
proportional to the Euler characteristic $\chi=2-2g$, and is true in higher dimensions as
well.  If such manifolds could be used to get compactifications with negative or small
positive vacuum energy, it seems almost inevitable that they would dominate the landscape.

\end{itemize}

In this work, we will give an explicit expression for the effective potential in
supergravity compactifications, \eq{Veff-final}, mostly answering the first two
questions, and making a start on the third.  

Besides understanding solutions, another important application of an effective potential is to study
time dependence.  In particular, slow-roll inflation can be described as gradient descent.
With this potential, this leads to equations similar to the Ricci flow equations, as we will
discuss elsewhere.  We might also hope that this will shed light on the deeper questions regarding
the existence of the effective potential raised by Banks in \cite{Banks:2004xh}, or on the old
problem of the conformal factor in Euclidean quantum gravity.

\section{Effective potential}
\label{s:general}

\subsection{General discussion}

We consider $D$-dimensional solutions of general relativity coupled to
matter, with an action such as
\be \label{eq:gr-action}
S = \int \sqrt{-g} \left( R^{(D)} - \half\sum_p |F^{(p)}|^2 \right).
\ee
We take metric signature $-+++\ldots$, and
define units so that the fundamental
Planck scale $M_{Pl,D}=1$.  

The equations of motion are the Einstein equation,
\be \label{eq:D-Einstein}
R_{MN} - \half g_{MN} R = T_{MN}
\ee
coupled to $p$-form gauge field strengths, with equation of motion
$d*F=0$ and stress-energy
\be \label{eq:flux-T}
T_{MN} = p F_{MI_1\cdots I_{p-1}} F_N^{I_1\cdots I_{p-1}} -  \half g_{MN} F^2 .
\ee
We choose a nonstandard normalization so that the case $p=0$ can be treated uniformly.

At least in the absence of Chern-Simons terms, one can choose to use an action
in terms of either $F^{(p)}$ or its dual $F^{(D-p)}$.  In the following,
we will use this freedom to consider all background fluxes as magnetic fluxes (so, $F_{0\ldots}=0$),
just to simplify the discussion.

We consider compactification on a $k=D-d$-dimensional compact manifold $M$
to $d$-dimensional maximally symmetric space-time (Minkowski, AdS, dS).  
Whenever there is any ambiguity, we superscript the metric and curvatures with
the dimension of space-time $D,k$ or $d$.  We superscript forms with their rank -- since
forms can be pulled back, generally there is no ambiguity.

We make a Kaluza-Klein warped metric ansatz, 
\begin{equation} \label{eq:simple-metric}
ds^2 = \eta_{\mu\nu} e^{2A(y)} dx^\mu dx^\nu + g_{ij}(y) dy^i dy^j .
\end{equation}
Here $\eta_{\mu\nu}$ is a maximally symmetric metric in $d$ dimensions, 
which could be dS, AdS or Minkowski, and
$g_{ij}$ is a metric on $M$.

Our goal is to write a $d$-dimensional effective action,\footnote{
We absorb the customary $1/16\pi$ into our definition of $G_N$.}
\be \label{eq:d-kk-action}
S^{(d)} = \int \sqrt{-g^{(d)}} \left( \frac{1}{G_N} 
 R^{(d)} - \half G_{ab}(\phi) \partial \phi^a \partial\phi^b - 2 V_{eff}(\phi) + \ldots\right),
\ee
whose equations of motion agree with the KK reduction of the
$D$-dimensional Einstein equations.  By $\ldots$ we denote gauge field actions and other
terms we will not treat in detail here.  Thus, the $d$-dimensional Einstein equation is
\be \label{eq:d-dim-eff-Einstein}
R^{(d)}_{\mu\nu} - \half g_{\mu\nu} R^{(d)} = G_N T^{(d)}_{\mu\nu}
\ee
with
\be
T^{(d)}_{\mu\nu} = -\half G_{ab}(\phi) \partial_\mu \phi^a \partial_\nu\phi^b 
- \half g_{\mu\nu} \left( \half G_{ab}(\phi) \partial \phi^a \partial\phi^b + 2 V_{eff}(\phi) \right) ;
\ee
in particular $T_{00}=V_{eff}+\ldots$.

In principle, we now want to rewrite the $D$-dimensional fields
using a mode decomposition, as appears in \cite{Duff:1986hr} and many other works.
We would then substitute these expressions into the $k$-dimensional part of the action,
do the integral over $X$ and reinterpret the results as terms in \eq{d-kk-action}.
Any term with no space-time derivatives would become part of the
effective potential.

The main difficulty in doing explicit mode expansions is to diagonalize the various
Laplacians on $X$ which appear as kinetic terms.
In a mathematical sense this step is well understood,
and is not directly relevant for our purposes.
Since we do not need to diagonalize the metric $G_{ab}(\phi)$ on field space,
any complete basis for functions on $X$ would suffice.
Thus, we can regard the $D$-dimensional fields as generating
functions for the totality of massive modes in the $d$-dimensional compactified theory.

\subsection{Einstein equations}
\label{ss:einstein-eq}

The subtleties we are concerned with appear elsewhere, and can be seen by reviewing
the standard discussion of compactification based on the $D$-dimensional Einstein equations.
Since these look rather different in $d$ and $k$ dimensions,
we consider the two components separately.   The $d$-dimensional components
can be written
\be
\label{eq:do-Einstein}
R^{(d)}_{\mu\nu} - \half g_{\mu\nu} R^{(d)} = T^{(d)}_{\mu\nu} + \half g_{\mu\nu} R^{(k)} 
\ee
and we would like to interpret the right-hand side as an effective 
$d$-dimensional stress tensor.
Since the $d$ dimensions have maximal symmetry, we lose nothing
by taking the trace,
\bea \label{eq:tr-d-Einstein}
-\frac{d-2}{2} R^{(d)} &=&  T^{(d)} + \frac{d}{2} R^{(k)} \\
 &=& -d\cdot \Lambda
\eea
The right hand side of this equation defines an effective cosmological constant,
\begin{equation}  \label{eq:eff-cc}
\Lambda = T_{00} -\half  R^{(k)} ,
\end{equation}
with a contribution from the curvature of $X$.  To get a
$d$-dimensional equation, we then integrate over $X$, 
producing \eq{k-R-pot}.
More precisely, all three terms depend on the volume of $X$, which
will lead to the dependence on Newton's constant $1/G_N$ in the result.

For standard (classical) sources of energy, except for negative potential energy (which is not
present in the string/M theory case),
$T_{00}\ge 0$, so naively one might expect de Sitter to be easy to realize.  On the other hand,
the scalar curvature also contributes, so one needs to decide which
effect is more important.

The $k$-dimensional curvature can be determined by using the
$D$-dimensional trace of the Einstein equation,
\be \label{eq:D-Einstein-tr}
\left(1-\frac{D}{2}\right) R = T^{(d)} + T^{(k)} ,
\ee
 to solve for $R$ in \eq{D-Einstein}, obtaining
\be \label{eq:k-Einstein}
R^{(k)}_{ij} = T^{(k)}_{ij} - \frac{1}{D-2} g_{ij} \left(T^{(d)} + T^{(k)}\right) .
\ee

Taking the trace, one sees that the sign of $R^{(k)}$
is correlated to that of $T^{(k)}$, while the $T^{(d)}=-dT_{00}$ contribution now favors
positive internal curvature.
Thus, the effects of $T^{(d)}$ in the two equations conflict with each other. 

Solving for $R^{(k)}$ and substituting back into \eq{eff-cc}, we get
\bea 
\Lambda &=& T_{00} -\half \left( \frac{d-2}{D-2} T^{(k)} - \frac{k}{D-2} T^{(d)} \right) \\
&=& -\frac{(d-2)(k-2)}{2(D-2)} T_{00} - \frac{d-2}{2(D-2)} T^{(k)} .
\label{eq:Lambda-eff}
\eea
Evidently the total effect of $T_{00}$ always favors AdS, with the only
hope for dS being to have $T^{(k)}$ very negative compared to $T_{00}$,
more precisely
\be \label{eq:T-flux-bound}
T^{(k)} < - (k-2) T_{00} .
\ee
Now, taking magnetic flux $F_{0\ldots}=0$ in \eq{flux-T}, we have
\be \label{eq:F-p-relation}
T^{(k)} = (p-\frac{k}{2}) F^2 = (2p-k) T_{00} .
\ee
Thus, this condition will never hold for any $p\ge 1$, or sums of such stress-tensors.

Thus, de Sitter space-time can only be realized if the matter stress tensor
violates the strong energy condition \cite{Wald:1984rg}, which is
\be
0 \le R_{00} = T_{00} - \frac{1}{D-2} g_{00} T  .
\ee
Terms which violate this condition include positive potential energy,
certain string/M theory corrections, and the $p=0$ flux which appears in massive IIa 
supergravity.  Conceptually, this is the statement that the strong energy condition 
is preserved under restriction; if it holds on $M\times X$ then it will hold on $M$
\cite{Gibbons-review,Gibbons:2003gb}.  This no-go theorem was independently
rediscovered by deWit-Hari Dass-Smit \cite{deWit} and Maldacena-Nunez
\cite{Maldacena-Nunez}.

\bigskip

Coming back to \eq{eff-cc}, while it 
might seem that we have justified \eq{k-R-pot},  there is an issue.
We needed to use the $D$-dimensional Einstein equation \eq{D-Einstein}
in order to determine $R^{(k)}$. 
On the other hand, the effective potential was supposed to be defined
independently of solving the $d$ or $D$-dimensional Einstein equation,
so it is not clear that \eq{k-R-pot} can be interpreted in this way.

Certainly, we cannot interpret \eq{Lambda-eff} as an effective potential.  This 
is clear from
the nontrivial factor in the relation between $T_{00}$ and $\Lambda$,
which even has the wrong sign.
This factor arose because we solved various equations of motion, including
\eq{D-Einstein}, and substituted the solutions.  While the final results are correct,  
identifying terms in partial results with terms in an effective action is usually not.  

We could correctly
identify the curvature term in \eq{eff-cc} with a term \eq{k-R-pot} in the
$d$-dimensional effective potential, if we could do this without solving any 
equations of motion.
However, we cannot simply relax \eq{D-Einstein}, as part of it (the $G_{00}$ component)
is in fact a constraint.  As we mentioned in the introduction, if we do not enforce this
constraint, the Einstein action and thus the potential \eq{k-R-pot} is unbounded below
and has no local minima.  To see this, evaluate \eq{k-R-pot} (using the results in the
appendix) for the conformally transformed metrics
\be
\tg_{ij} = e^{2B} g_{ij} ,
\ee
obtaining
\be
V_{eff} = \half \int \sqrt{g} e^{(k-2)B} \left( -R^{(k)}[g] -(k-2) (k-1) (\nabla B)^2 \right) ,
\ee
with the wrong sign for the derivative term.

Now it is true that the potential in gravity and supergravity is often unbounded below,
but only to one or a few directions in field space.  On the other hand, with
a negative definite spatial derivative term, every 
short wavelength perturbation of the conformal factor is tachyonic, so the theory
would be completely unstable.
Somehow, this problem must be fixed by incorporating the constraints of the $D$-dimensional theory.

Having seen the problem, it is not hard to come up with a strategy for dealing with it.
The constraints have two parts, a ``zero mode'' part
and a ``non-zero mode'' part.  The former become constraints in the
$d$-dimensional theory, while the latter need to be solved in the process of dimensional
reduction.

\subsection{Incorporating the warp factor}

We now study the dependence of the action on the warp factor, 
and consider the general warped ansatz 
\begin{equation} \label{eq:metric}
ds^2 \equiv \tg_{AB} dx^A dx^B = e^{2A(y)}\eta_{\mu\nu} dx^\mu dx^\nu +
 e^{2B(y)} g_{ij} dy^i dy^j .
\end{equation}
We denote the metric with $A=B=0$ as $g_{AB}$, and the
metric \eq{metric} as $\tg_{AB}$.  

As $B$ is redundant with $g_{ij}$, for now
we will take $B=0$, until we are ready to discuss 
this part of the problem.   
Using the results in the appendix,
the $k$-dimensional part of the
Einstein equations becomes
\be \label{eq:k-Einstein-A}
R^{(k)}_{ij} - \half g_{ij} R^{(k)} = T^{(k)}_{ij} + d \nabla_i\nabla_j A + d \nabla_i A \nabla_j A
 + g_{ij} \left( \half e^{-2A} R^{(d)} -d \nabla^2 A - \frac{d(d+1)}{2} (\nabla A)^2 .
 \right) 
\ee

The $d$-dimensional trace \eq{tr-d-Einstein} becomes
\be \label{eq:d-trace}
-\frac{d-2}{2}e^{-2A} R^{(d)} +
d(d-1) \nabla^2 A + \frac{d^2(d-1)}{2} (\nabla A)^2 = T^{(d)} + \frac{d}{2}R^{(k)}.
\label{eq:d-constraint}
\ee
with
\be
T^{(d)} = - \frac{d}{2}\sum_p |F^{(p)}|^2 + T^{(d)}_{string}
\ee
At this point we have added the term $T^{(d)}_{string}$ which represents 
the non-classical sources present in superstring theory.  We will not use its
detailed form, only the fact that it allows violating the inequality $T^{(d)} \le 0$.

Note that, except for $R^{(d)}$, every term in \eq{d-trace} comes with the same
weight $e^{\alpha A}$ with $\alpha=0$.\footnote{
See for example equations (2.12) and (2.13) of \cite{deWit}.}
This dependence has only
two sources: the overall $\sqrt{g^{(d)}}$, which sits in front of every term, and
explicit factors of the $d$-dimensional metric $g_{\mu\nu}$ or its inverse.
However, since every term is a scalar in space-time, and 
does not contain $d$-dimensional derivatives, the $d$-dimensional metric
cannot appear.  This also includes sources in $T^{(d)}_{string}$ which are space-time scalars.
While the argument we just gave does not cover Chern-Simons terms or electric field 
strengths, they also have the same weight, as we argue in \S \ref{ss:dilaton}.

\eq{d-trace} can be dramatically simplified by the change of variable
\be \label{eq:def-u}
u \equiv e^{dA/2} ,
\ee
and becomes
\be \label{eq:u-constraint}
-\frac{d-2}{2}R^{(d)} u^{1-4/d} = -{2(d-1)} \nabla^2 u +\left(\frac{d}{2}R^{(k)}+T^{(d)}\right)u .
\ee
This change of variables is used to great advantage in related mathematical
work, on the Yamabe problem \cite{yamabe} and in Perelman's entropy functional \cite{perelman}.
Readers who have looked at this, or at Tseytlin's discussion of an entropy functional
for sigma models \cite{Tseytlin:2006ak}
will recognize many ingredients of the following discussion, for example \eq{V-as-eigenvalue}.

\subsection{Effective potential}
\label{ss:effpot}

We seek a functional of $u$ and $g_{ij}$ (and matter fields) whose variation produces
the two Einstein equations.  Let us try direct substitution of the ansatz into the
$D$-dimensional action.
We continue to take $B=0$, then
\bea \label{eq:action}
S_{eff,R} &=& \int \sqrt{\tg}\left(\tR^{(D)}-\half |F^{(p)}|^2\right) \\ \nonumber
 &=& \int \sqrt{g} \left( u^{2-4/d}R^{(d)} +  u^2 
R^{(k)} +\frac{4(d-1)}{d} (\nabla u)^2 -\frac{u^2}{2} |F^{(p)}|^2\right) .
\eea
The similarity to the string action with the dilaton is not coincidental, and is
because one can also obtain the dilaton by dimensional reduction.

Varying this with respect to $g_{ij}^{(k)}$, we get 
\bea \label{eq:k-einstein-u}
u^2 \left(R^{(k)}_{ij} - \half g_{ij} R^{(k)} \right) - \nabla_i \nabla_j u^2
+ g_{ij} \nabla^2 u^2 &=& u^2 T^{(k)}_{ij} - \frac{4(d-1)}{d} \nabla_i u \nabla_j u \\
\nonumber
 &&- \half g_{ij} \left(  - \frac{4(d-1)}{d} (\nabla u)^2- R^{(d)}  u^{2-4/d}\right) 
\eea
which can be checked to be equivalent to
\eq{k-Einstein-A} with the substitution \eq{def-u}.  Here $T^{(k)}_{ij}$ is defined to be
the variation of the matter action with respect to $g_{ij}^{(k)}$.

Since $e^{2A}=u^{4/d}$, the variation
$\delta g^{\mu\nu}=g^{\mu\nu}$ which leads to \eq{d-trace}, should be equivalent
to varying with respect to $u$ and multiplying by $-d/4$, and it is.

If we now try to identify the effective potential by direct comparison of \eq{gr-action} and
\eq{d-kk-action}, we find
\be \label{eq:Veff-R}
\int \sqrt{g}  \frac{u^2}{2} \left(- R^{(k)} +\half \sum_p |F^{(p)}|^2 - \frac{2}{d}T^{(d)}_{string}\right)
- \frac{2(d-1)}{d} (\nabla u)^2 
\ee
which is close but not quite right, because
the $R^{(d)}$ term is missing from the equations
of motion.  However, we cannot simply add it back,
as we are trying to define an effective potential which is purely a functional of the
$k$-dimensional fields.

Rather, we implement the strategy described at the end of \S \ref{ss:einstein-eq}.
The equation $0=\delta S/\delta u$ is a constraint equation, so we need to enforce
it in the definition of $V_{eff}$, except for a zero mode part.  The zero mode part is exactly
the part sourced by the $R^{(d)}$ term, and thus we can do this by replacing $R^{(d)}$ with
an undetermined constant $C$.  If the $d$-dimensional equations equate this to $R^{(d)}$,
we will get the correct $k$-dimensional equations of motion.

On the other hand, from the point of view of the $d$-dimensional equations
\eq{d-dim-eff-Einstein}, we are adding an extra term to $V_{eff}$.  To reconcile the
various equations, we need to subtract the same term without the $u$ dependence.

While this may seem a bit {\it ad hoc}, there is another way to justify it.  The term in
\eq{action} which becomes the $d$-dimensional Einstein term is
\be \label{eq:dd-Einstein}
\int \sqrt{g} u^{2-4/d}\, R^{(d)} .
\ee
Thus we identify the $d$-dimensional Newton's constant as
\be \label{eq:def-GN-one}
\frac{1}{G_N} = M_{Planck,d}^{d-2} = \int \sqrt{g} e^{(d-2)A} = \int \sqrt{g} u^{2-4/d} ,
\ee
so we can interpret the constant $C$ as a Lagrange multiplier which enforces
this definition.
This leads to
\be \label{eq:Veff}
V_{eff} =  \half\int \sqrt{g} \left[
 u^2 \left(- R^{(k)} +\half \sum_p |F^{(p)}|^2 - \frac{2}{d}T^{(d)}_{string}\right)
- \frac{4(d-1)}{d} (\nabla u)^2 \right]+ \half C \left(\frac{1}{G_N} - \int\sqrt{g} u^{2-4/d}\right) 
\ee

With this definition, the $\delta V_{eff}/\delta u=0$ constraint becomes
\be \label{eq:C-constraint}
-\frac{d-2}{2}C u^{1-4/d} = -{2(d-1)} \nabla^2 u +\left(T^{(d)} + \frac{d}{2}R^{(k)}\right)u ,
\ee
Given $C$, this is a sensible constraint, but we need to
explain how to choose $C$, and how this eventually implies \eq{u-constraint}.
Before explaining this, let us look at how we would solve \eq{C-constraint}.

\subsection{Solving the constraint}

To begin, let us set $C=0$, and instead add
a term $\lambda u$ on the left hand side.  We get a Schr\"odinger-type
equation on $X$,\footnote{
This equation (with $T^{(d)}=0$) appears in \cite{perelman,Tseytlin:2006ak},
although not with the interpretation of $u$ as a warp factor.
It is also the $d\rightarrow \infty$ limit of \eq{C-constraint}, a relation used
in the math literature.  Also, on replacing $d$ with $2-k$, 
\eq{C-constraint} with $T^{(d)}=0$ becomes the Yamabe equation \cite{yamabe}.}
\be \label{eq:schrodinger}
\lambda u = - \nabla^2 u +\frac{d}{2(d-1)}U\cdot u .
\ee
with a ``local potential'' 
\be \label{eq:local-potential}
U \equiv \half R^{(k)} + \frac{1}{d}T^{(d)} .
\ee
As is very familiar, 
such an equation has solutions $u_i$ for a discrete spectrum of eigenvalues $\lambda_i$,
and the $u_i$ form a complete basis for functions on $X$.

This is the relevant equation for Minkowski space-time
with $R^{(d)}=0$, so let us discuss this case first.
Since $u>0$ by its definition \eq{def-u}, 
the only acceptable solution is the ground state $u_0$.  However,
the eigenvalue $\lambda_0$
typically will not be zero.  For this to be true, the integral of the potential (against $u$) must be zero,
which is the ``warp factor constraint''  of \cite{Becker:1996gj,Giddings:2001yu}.
For compactification with $X$ Ricci flat
and nonzero flux, one can only satisfy this by adding sources with $T^{(d)}_{string}>0$, 
such as orientifolds or certain $\alpha'$ corrections.
The same will be true if we try to realize Minkowski space-time using an
$X$ with negative total scalar curvature $\int \sqrt{g}R^{(k)}<0$.  

Keeping in mind our Schr\"odinger
equation intuition, we return to the actual constraint equation \eq{C-constraint}.
Integrating over $X$, we derive the warp factor constraint,
\be \label{eq:wf-con}
-\frac{d-2}{2}C \int \sqrt{g} u^{1-4/d} = \int \sqrt{g} u \left(T^{(d)} + \frac{d}{2}R^{(k)}\right) .
\ee
While this is still a necessary condition, once we allow $C\ne 0$
it can always be solved, and relates $C$ to the scale of $u$.  

This is not yet the constraint of the no-go theorems \cite{deWit,Maldacena-Nunez}, 
which (as we discuss in \S \ref{ss:k-dim}) 
uses the $k$-dimensional Einstein equation to control the
sign of the right-hand-side.  If one knows that the local potential \eq{local-potential} has a definite
sign, clearly $C$ (and ultimately $R^{(d)}$) must have the opposite sign.  However,
the local potential $U$ need not have a definite sign, in which case one needs to know 
the warp factor $u$
to find the sign of the integral in \eq{wf-con}.
Even when it has a definite sign, 
we need this information to estimate $C$.

Rather fortuitously, for $d=4$, the equation \eq{C-constraint} is linear inhomogeneous, so we can
easily get more information.  Given the normalized eigenfunctions $u_i$, satisfying
\eq{schrodinger} with eigenvalues $\lambda_i$ and
\be
\int \sqrt{g} u_i u_j = \delta_{ij} ,
\ee
we can write the solution as
\be \label{eq:gen-solution}
u(y) = -\frac{1}{6}C\sum_i u_i(y) \frac{1}{\lambda_i}\int_X \sqrt{g}\, u_i .
\ee
If there are any modes with $\lambda_i=0$ (normally there will not be), they can be left out of the sum,
and enter the solution with undetermined coefficients.

To summarize, although we cannot solve the constraint \eq{u-constraint}
before knowing the $d=4$-dimensional
curvature $R^{(d)}$, we can solve it up to an overall coefficient.  The equation
\eq{C-constraint}
obtained by varying \eq{Veff} is equivalent to \eq{u-constraint}; we just need to relate
the coefficient $C$ to $R^{(d)}$.

The same idea can be applied in $d\ne 4$, and by rescaling $u$ one can 
set $C=1$ in \eq{C-constraint}, restoring it with the relation
$u \propto C^{d/4}$.   Of course, one would have to solve a nonlinear equation.
At first sight, the cases $d>4$ would appear similar to $d=4$, as the nonlinearity
is mild.  On the other hand, for $d=3$ the source term blows up as $u\rightarrow 0$,
which looks significant; however we leave the analysis of this for subsequent work.

\subsection{Interpreting the constraint}
\label{ss:int-constraint}

To summarize the discussion so far, we seek a $k$-dimensional functional whose
variation leads to the Einstein equations for maximally symmetric compactifications.
By direct substitution of the ansatz, we obtain \eq{action} which has this property,
but it is not purely $k$-dimensional, since 
it depends on the $d$-dimensional curvature $R^{(d)}$.

Usually (this is a convention), the $d$-dimensional effective action is defined in
Einstein frame, {\it i.e.} with
fixed Newton's constant $G_N$.    But this is easy to obtain from \eq{action}, because
$R^{(d)}$ and $G_N$ are conjugate variables.  Thus the effective potential is the
Legendre transform of \eq{action}, 
\be
V_{eff} =  \frac{1}{2 G_N} R^{(d)} - \half S_{eff,R}\,\bigg|_{\delta V_{eff}/\delta R^{(d)}=0} \\
\ee
with $R^{(d)}$ relabeled $C$, and the factor $\half$ introduced to agree with the conventions
of \eq{d-kk-action}.  This reproduces \eq{Veff}.

The condition $\delta V_{eff}/\delta R^{(d)}=0$ defining the Newton constant is
\be \label{eq:def-GN}
\frac{1}{G_N} = M_{Planck,d}^{d-2} = \int \sqrt{g} e^{(d-2)A} = \int \sqrt{g} u^{2-4/d} .
\ee
This is an independent  condition on $u$ and can also be used to determine
the overall normalization of the warp factor.  Thus, we choose the coefficient $C$
in order to satisfy this condition. 


Substituting the solution of the constraint \eq{C-constraint} into
the effective potential $V_{eff}$ defined in \eq{Veff}, and using the definition \eq{def-GN},
we get
\be \label{eq:C-proportion}
V_{eff} = \frac{d-2}{2d}  \frac{C}{G_N}   .
\ee
Thus, if we use this effective potential in 
\eq{d-kk-action}, and impose the Einstein equation following from this action,
we obtain \eq{tr-d-Einstein} with $C=R^{(d)}$, and have reproduced the discussion
of \S \ref{s:general}.  We stress that we do not need to know $R^{(d)}$ to compute 
it, rather this is done by solving \eq{C-constraint}, and then imposing \eq{def-GN}.
Nor does one need to impose the $k$-dimensional Einstein equations; it is defined
for general metric and field configurations.

Substituting \eq{gen-solution} into \eq{def-GN} (for $d=4$), we find that 
\be
\frac{1}{C} = -\frac{1}{6}G_N
\sum_i \frac{1}{\lambda_i} \bigg|\int \sqrt{g} u_i\bigg|^2 .
\ee
and
\be \label{eq:eigen-Veff}
\frac{1}{G_N^2 V_{eff}} = 
-\frac{2}{3}\sum_i \frac{1}{\lambda_i} \bigg|\int \sqrt{g} u_i\bigg|^2 ,
\ee
an explicit expression for $V_{eff}$ in terms of $k$-dimensional quantities.

\subsection{Incorporating the dilaton and other fields}
\label{ss:dilaton}

Essentially the same results apply to very general matter theories coupled
to Einstein gravity; in particular supergravity, because
every term in the effective potential will have the
same dependence on the warp factor.   A general argument to this effect is as follows.
Since the effective potential is a scalar, whose integral makes a contribution to the  
$d$-dimensional action proportional to the space-time volume,
the same factor of the $d$-dimensional space-time volume element must appear in every term.
Then, using the definition
made in \eq{simple-metric}, the dependence on the warp factor is the same
as the dependence on this volume element.  

For example, the
NS sector of the type II string actions in string frame leads to
\bea \nonumber
V_{eff} &=&  \int \sqrt{g^{(k)}} e^{-2\Phi} \left[  u^2 \left(-R^{(k)} - 4(\nabla\Phi)^2 + \half 
 |H^{(3)}|^2 - T^{(d)}_{string}  \right) 
- \frac{4(d-1)}{d} (\nabla u)^2 \right] \\
&&
+ C  \left( \frac{1}{G_N} - \int \sqrt{g} e^{-2\Phi} u^{2-4/d} \right).  \label{eq:Veff-typeII}
\eea
In this form, it is not manifest that the dilaton contribution to the potential is bounded
below.  One could make this manifest either by going to Einstein frame in $k$ dimensions, or
(for IIa theory) going to an M theory description in $k+1$ dimensions.  Our later arguments
that the potential is bounded below will assume that this has been done.

Two cases in which the warp factor dependence may not be immediately 
obvious are contributions from electric flux, and contributions from the
Chern-Simons terms.  The easy way to argue in the first case is to use duality to
reexpress the action in terms of magnetic flux.  Thus, a $p$-form electric flux
behaves in expressions such as \eq{F-p-relation} like a $D-p$-form magnetic flux.

A more direct argument to this effect uses the fact that a background electric flux compatible
with maximal space-time symmetry must have $p\ge d$, and transforms like the 
$d$-dimensional volume form multiplied by a $p-d$-form on $X$.  Then, it is
quantized in units of the $d$-dimensional volume form 
\cite{Bousso:2000xa}.  Taking into account the space-time metric in the term $|F^{(p)}|^2$,
this term is independent of the warp factor, so again the entire dependence comes from
the volume form $\sqrt{g^{(d)}}$.

As for the Chern-Simons terms,
these will only contribute to the effective potential
in the presence of a background electric field; for example in M theory we can have
\be
\int C^{(3)} \wedge G^{(4)} \wedge G^{(4)}
\ee
and a background $G^{(4)}=G dx^0 dx^1 dx^2 dx^3$ in $d=4$.
Again, the quantization condition on the electric field will force
the same warp factor dependence.

\section{Physical discussion}
\label{s:physical}

Let us recap  the final expression for the effective potential.  We have eliminated
all dependence on $d$-dimensional physics except the number $d$, which we now set to $d=4$.
The metric, curvature, fluxes and warp factor $u$ are all defined on $X$, as is the additional
``stringy source'' $T^{(d)}_{string}$.
\be \label{eq:Veff-final}
V_{eff} =  \half\int \sqrt{g}  \left[ u^2 \left(-R + \half 
\sum_p |F^{(p)}|^2 - \half T^{(d)}_{string}\right) 
- 3 (\nabla u)^2 \right]
+ \half C  \left(\frac{1}{G_N}-\int \sqrt{g}\,u\right) .
\ee
To use it, one enforces the warp factor constraint $\delta V_{eff}/\delta u=0$, 
which in $d=4$ is linear,
\be \label{eq:wfc}
-\frac{1}{6}C  = \left(- \nabla^2 + \frac{2}{3}U\right)u ; \qquad 
 U = \half R - \frac{1}{4}\sum_p |F^{(p)}|^2 + \frac{1}{4}T^{(d)}_{string} 
\ee
thus determining $C$ and $u$ up to an overall normalization, and the
warped volume constraint 
$$
\frac{1}{G_N} =  \int \sqrt{g} u ,
$$
thus determining the normalization.  Due to the simple form of \eq{Veff-final},
its final value on substituting $u$ is simply  $C/4G_N$, as in \eq{C-proportion}.

\subsection{Physical regimes}

With the main result in hand, let us discuss some of its physics in $d=4$.
First, we consider a constant warp factor, 
\be
u=\frac{1}{G_N \Vol X } ,
\ee 
so \eq{Veff-final} reduces to
\be \label{eq:Veff-unwarped}
V_{eff,unwarped} = - \frac{\int\sqrt{g}\, U}{(G_N \Vol X)^2}.
\ee
In this case, the original intuition leading to \eq{k-R-pot} is correct.  This
will be exact if $U$ is constant.

Before we continue,
there are two general conventions we could take for $G_N$.  When discussing
fixing of moduli, such as $\Vol X$, we want to exhibit the dependence of different terms
of the potential on the moduli, so we should choose a fixed $G_N$ in fundamental units.
In this case we have a universal $1/(\Vol X)^2$
(in $d=4$) prefactor for the potential, as in \cite{Giddings:2003zw}.  

As the simplest possible example, taking $X=S^k$, with $\Vol X=L^k$,
and turning on flux $F^{(p)}$ with $p=k$,
we have
\bea
V_{eff} &=& \half\int \sqrt{g} u^2 (-R + \half ||F||^2) \\
 &=& \frac{1}{2 G_N^2 L^{2k}} \left( -R\,L^{k-2} + \frac{F^2}{2L^{k}}\right)
\eea
where $R>0$ is the curvature of a diameter $1$ sphere.
This potential has a unique negative minimum.  In this approximation, one recovers many
of the effective potentials found in the string compactification literature.

Once the volume is fixed, one is better off taking $G_N \propto 1/ \Vol X$, in which
case $u \sim 1$.  This is the convention we will follow below.

As we now explain, 
there appear to be two general regimes, distinguished by whether the warp factor
is slowly varying or not.
Since the new physics of the warp factor has to do with its variation,
the first might also be called ``weak warping,'' or some more euphonious version of this.

\subsection{Slowly varying warp factor}

If the variation of $U$ is small,  we can try to
ignore the derivatives in \eq{wfc}, to find
\be \label{eq:no-deriv-approx}
u = -\frac{C}{4U} =  -\frac{G_N V_{eff}}{U} .
\ee
One necessary condition for this to make sense is that 
$U$ must have a definite sign.  
Treating the derivatives as a perturbation, and solving to first order, another
condition for this approximation to be good is
\be \label{eq:slow-warping}
\bigg|\nabla^2\left(\frac{1}{U}\right)\bigg| << 1.
\ee
While these conditions can be satisfied, for example in AdS compactifications,
they are quite restrictive.  Localized sources like branes and orientifold planes
will violate them.  Even without these, if
$U$ varies on a scale $L$ (in fundamental units), say set by the size of $X$, then we require
\be \label{eq:slow-warp-is-good}
L^{k+2} \cdot |G_N^2 V_{eff}|  >> 1 .
\ee
Thus one requires large vacuum energy, large volume or both.

In this approximation, $V_{eff}$ is always smaller in magnitude than
\eq{Veff-unwarped}.  To see this, we begin by substituting \eq{no-deriv-approx}
into \eq{Veff-final}, to get an explicit expression for $V_{eff}$ in terms of $U$,
\be \label{eq:Veff-slow-warped}
\frac{1}{ G_N^2 V_{eff} }= \pm\norm{ \frac{1}{U} }_{L^{2-4/d}} ,
\ee
where the sign is opposite to that of $U$, and we use the mathematical notation
\be
\norm{ f }_{L^\alpha} = \left(\int \sqrt{g}\, |f|^\alpha \right)^{1/\alpha} .
\ee
Then, we use
the Cauchy-Schwarz inequality
\be
\norm{f}\, \norm{ g } \ge (f,g)^2 ,
\ee
applied to $f=U^{1-2/d}$ and $g=U^{2/d-1}$.

Thus, in this regime warping increases the effective potential for AdS, and
would decrease it for dS.

\subsection{The ground state approximation}
\label{s:ground-state}

When the condition \eq{slow-warping} is not satisfied, we need to take the
derivatives into account.  
Let us consider the idea that, 
to a good approximation, the warp factor will be proportional to
the ground state wave function $u_0$.
When it is, we can drop the other terms in \eq{gen-solution}, to find
\be \label{eq:truncate-ground-state}
u \sim c u_0; \qquad \frac{1}{c} \equiv  G_N\int\sqrt{g} u_0; \qquad
C \sim -6G_Nc^2\lambda_0
\ee
and
\be \label{eq:V-as-eigenvalue}
V_{eff}=-\frac{3}{2}c^2 \lambda_0.
\ee
Here $u_0\sim 1/\sqrt{\Vol X}$ (since it is normalized), so $c\sim \sqrt{\Vol X}$
and $u \sim 1$.

Of course, a constant warp factor is the ground state for constant $U$, so
there will some nearby regime in which this is a good approximation.    
Here are various arguments that this regime is large:
\begin{itemize}
\item 
Out of all the source terms  $\int\sqrt{g} u_i$ in \eq{gen-solution}, only $i=0$ has a
positive definite integrand.  
For the other eigenmodes, the integral will have cancellations, typically making it
much smaller than $1/G_N$.
\item Since the other eigenfunctions always have
nodes and are negative in some part of $X$, if they come in with significant coefficients,
the solution is in danger of violating the
consistency condition that $u>0$ at each point.
\item
Normally, the $1/\lambda_i$ factor will be maximized for $i=0$.
This is clear if the spectrum is positive.  If $\lambda_0<0$, there are good reasons
(which we discuss below) to expect that the other $\lambda_i>0$, in which case all
but a few will have $\lambda_i>>|\lambda_0|$.
\end{itemize}

In \S \ref{ss:lambda-zero}, we will further discuss the validity of this approximation.  
To oversimplify a bit, it will be good
in the opposite regime from \eq{slow-warp-is-good},
\be \label{eq:ground-state-is-good}
L^{k+2} \cdot |G_N^2 V_{eff}|  \lesssim 1 .
\ee

In this approximation,
warping significantly increases the effective potential:
\be
V_{eff} = -\frac{3}{2}c^2 \lambda_0 > V_{eff,unwarped} = -\frac{\int\sqrt{g}\, U}{(G_N \Vol X)^2}.
\ee
This follows from 
the variational bound for the wave function $u={\rm constant}$, 
\be
-\lambda_0 \ge -\frac{\int \sqrt{g}\, U}{\Vol X} ,
\ee
the Cauchy-Schwarz inequality
\be
\frac{1}{c^2} = G_N^2(\int \sqrt{g} u_0)^2) \le 
G_N^2\Vol X ,
\ee
and finally $3/2 > 1$.

\subsection{Basic picture}

The two approximations we discussed do not cover all the possibilities, and one might
study others, or develop a mixed picture by using each in different regions and 
patching them together.  However the general picture is similar enough in both that we leave this
for subsequent work, and instead try to outline the physical picture.

As is familiar from quantum mechanics, 
the ground state wave function will have most of its support in
potential wells.  This is also true (although less so) for \eq{no-deriv-approx} with $U>0$.

Here, the local potential is minus the energy density $T_{00}$, so 
the warp factor will be concentrated in regions of positive energy.  Conversely,
regions with negative energy will have this energy warped (or redshifted) away.

Most sources, such as fluxes, make positive contributions to the energy density.  This
also includes negative curvature.  On the other hand, positive curvature, and some
stringy sources like orientifolds, make negative contributions to the energy density.

Physically, the warp factor relates energy scales on $X$ to those in space-time.
In the language of AdS/CFT (see for example \cite{Verlinde:1999fy}),
one says that $u$ is large in the ``UV'' and small in the ``IR.''  The usual AdS solutions
are supported by $X$ with positive curvature.  This contributes a negative
energy density which outweighs the flux energy, and
sends $u\rightarrow 0$, consistent
with the picture we just gave.  Conversely, the parts of $X$ with positive energy density,
either from flux or from negative curvature, are the UV.

Thus, the basic physics is that, while the vacuum energy density is a sum of effects
from different regions of the manifold, each given by minus the local potential $U$,
we must take into account the local warp factor in adding energy densities, which we do
with the weighing $u^2 \cdot U$.
This favors the UV region, with positive energy, and thus raises $V_{eff}$.

This gives us a physical answer to the puzzle raised in the introduction, that varying
the conformal factor might send the effective potential to arbitrarily negative values.
Such variations produce positive scalar curvature, so decrease the warp factor, and
redshift away the negative energy.  In \S \ref{s:stability} we will try to demonstrate
this from the equations.

In an actual vacuum,
the local potential $U$ is determined by the $k$-dimensional Einstein equation
in a way we will discuss below.  If one can find solutions in which it
is negative in some regions and positive in others, since the warp
factor weighs the negative regions more heavily,
taking it into account should make it {\bf easier} to find de Sitter solutions.
On the other hand, the no-go theorems show that negative $U$ is not so easy to
obtain.

\subsection{Going through $\Lambda=0$}
\label{ss:lambda-zero}

What happens if one varies parameters in $k$ dimensions (say pseudo-moduli
or fluxes), so that $V_{eff}$ crosses zero?
While in conventional field theory, there is nothing special about zero energy, in gravity
there definitely is.  Can we see any sign of this in $k$ dimensions?  

As we recalled in the introduction, the constructions of de Sitter vacua using the effective potential,
implicitly or explicitly rely on precisely such a continuation of parameters.  For example, one
style of argument \cite{Kachru:2003aw} is to combine sources of vacuum energy which are computable
in a supersymmetric AdS vacuum, with a single additional supersymmetry breaking
energy which ``uplifts'' the vacuum to dS.  This might be justified by continuing the
parameters from a better controlled nonsupersymmetric AdS vacuum, or conversely some flaw
in the reasoning might appear at this step.  Thus, it is crucial to understand this point.

At first sight, the series solution \eq{gen-solution} for the warp factor looks singular
as an eigenvalue $\lambda$ passes through zero.  We can get a simple model for this
by considering the family of local potentials obtained by a simple shift $a$ of the energy,
\be \label{eq:U-shift-a}
U_a \equiv U + a .
\ee
In massive IIa theory, the $(F^{(0)})^2$ source has precisely this effect (with $a<0$).

Clearly the eigenfunctions $u_i$ are independent of $a$, while the eigenvalues simply
become $\lambda_i+a$.  Thus the series solution becomes
\be \label{eq:gen-solution-a}
u(y) = -\frac{C}{6}\sum_i u_i(y) \frac{1}{\lambda_i+a}\int \sqrt{g} u_i .
\ee
As we approach the resonance, it should be a good approximation to keep only
the resonant term in the sum.  Now, there are two cases:
\begin{itemize}
\item An excited state, {\it i.e.} $i>0$, passes through zero, $\lambda_i+a \sim 0$.
Assuming that the matrix element $\int\sqrt{g} u_i\ne 0$, which (in the absence of
some special consideration such as symmetry) will almost surely be true, we will
have $u \sim u_i$ near the resonance.  But
since the excited state wave functions $u_i$ all have nodes, this is inconsistent with
positivity of $u$.  Before we reach this point, we will find $u=0$ somewhere on $X$,
so our assumptions must break down.
\item On the other hand, if the ground state energy passes through zero, we are 
simply doing the same truncation to the ground state that we discussed in 
\S \ref{s:ground-state}.  All of the approximate results given there, such as
\eq{truncate-ground-state} and
\eq{V-as-eigenvalue}, have sensible continuations through $\lambda_0=0$.
\end{itemize}

Thus, we see no obstacle to continuing $V_{eff}$ through zero to a de Sitter vacuum with
$\lambda_0<0$, up to the point where $\lambda_1$ crosses zero.  Somewhere before
that point,  at a vacuum energy
\be
\Lambda \sim M_{Planck,4}^4 (\lambda_1-\lambda_0) ,
\ee
there will be some sort of transition.  It is not clear whether this can be
described using supergravity.  If it can, perhaps the
factorized form \eq{simple-metric} of the metric breaks down.  If not, then assuming $u$
is analytic (as will be true for $U$ analytic) and vanishes at a point $y_0$, the metric
near this point will look like (in $d=4$)
\be
ds^2 = \alpha |y-y_0| \eta_{\mu\nu} dx^\mu dx^\nu + dy^2 + \ldots ,
\ee
in other words an event horizon.  The precise meaning of this depends on the
metric on $X$, and since the $k$-dimensional Einstein equation has source
terms which blow up as $u\rightarrow 0$, we need to look at details of
the example to work this out.

One could get good estimates for this gap using general results on Schr\"odinger operators.
For example, if the wavefunction is localized to a potential well $U\sim m^2x^2/2$, 
there will be excited states with $\Delta\lambda\sim m$.  Again, the best way to do this
probably depends on details of the example at hand.

A simplified picture is to grant that
\be
\lambda_1 - \lambda_0 \sim L^{-2} ,
\ee
which leads to the bound \eq{ground-state-is-good}.

\subsection{Cosmological application}

In subsequent work we will try to develop a picture of early cosmological dynamics based on these
results.
The basic idea is that the equations of motion for slow roll inflation,
\be
3H \frac{\partial \phi^a}{\partial t} = - G^{ab}\frac{\partial V_{eff}}{\partial \phi^b} ,
\ee
can be computed using \eq{Veff-final}, to get a flow on the space of metrics and other
fields on $X$ similar to Ricci flow.  This is somewhat like Perelman's treatment of Ricci
flow as a gradient flow \cite{perelman}.

As is manifest from its definition, the effective potential $V_{eff}$ decreases under the
flow; thus one expects that the local potential $U$ becomes more positive, and the eigenvalues
$\lambda_i$ each increase, in a way we can roughly model by taking \eq{U-shift-a} with
$a \sim t$.
In this situation the consistency condition we just discussed might be a significant constraint
on the compactifications which lead to a sensible cosmology.

\section{Stability}
\label{s:stability}

As we saw in \S \ref{s:physical}, we expect the problem 
raised in the introduction, that the effective potential is unbounded below under
varying the conformal factor, to be solved by the warp factor.
Such variations produce positive scalar curvature, so decrease the warp factor, and
redshift away the negative energy.  

In this section we look at how one might show this from the equations.   In fact, at this
point we have a precise mathematical
\begin{conjecture}
Consider a conformal class of metrics 
$\tilde g = e^{2B} g$ on a $k$-dimensional manifold $X$; then the
functional \eq{Veff-final} evaluated at its critical point $\delta V/\delta u=0$, 
considered as a function on the space of all conformal factors $B$ with 
fixed warped volume and volume
\be
\int \sqrt{\tilde g} u^{2-4/d} = C_1; \qquad \int \sqrt{\tilde g} = C_2;
\ee
and all $F$, is bounded below. 
\end{conjecture}
To be precise, the volumes $C_1$ and $C_2$ should be
defined in the $k$-dimensional Einstein frame.\footnote{
By ``Einstein frame,'' one means conventions in which 
$V_{eff} \sim -\int \sqrt{g}R$
without any field-dependent prefactor, of the type which appears (for example) in 
\eq{Veff-typeII}.  Such a prefactor can be removed by a field-dependent
conformal transformation.}

Although this is not the same claim as the positive energy
theorem (there are AdS vacua), clearly it is related, and perhaps
it can proven using some variation of the arguments used there
\cite{Schoen:1979zz,Witten:1981mf}.  Here we make some nonrigorous but suggestive arguments.

\subsection{Linearized stability}

This was already checked for various explicit solutions in the early Kaluza-Klein
literature.  Let us check it at short distances, for which one can simply take a 
flat background metric.  Thus,
we allow a general conformal factor $B$ as in \eq{metric},
turning \eq{action} into
\bea
V_{eff} &=& \int 
-\sqrt{g}  e^{dA+(k-2)B}
 \left(R^{(k)} -2(k-1)\nabla^2 B - (k-2)(k-1)(\nabla B)^2+ d(d-1) (\nabla A)^2 \right) \nonumber\\
&=& -\int \sqrt{g}  e^{dA+(k-2)B}
 \left(R^{(k)} +2d(k-1)\nabla A\nabla B + (k-2)(k-1)(\nabla B)^2+ d(d-1) (\nabla A)^2 \right) 
 \nonumber\\
\eea
While every term individually is negative definite,
the determinant of the $2\times 2$ matrix of kinetic terms is
\be
\det = (k-1)(k-2)d(d-1) - d^2(k-1)^2 =  -d(k-1)(D-2) < 0
\ee
so it has signature $(+1,-1)$. 
As we expect, the conformal factor is the only mode with a wrong sign kinetic term.
Solving the constraint and substituting back in, can in principle produce a positive
definite effective potential.  We check that this is true at the linearized level
for short wavelength fluctuations of $B$; the linearization of \eq{d-constraint} is
$$
p^2 A = -\nabla^2 A =  \frac{k-1}{d-1} \nabla^2 B
$$
and the prefactor of the kinetic term becomes
\be \label{eq:prefactor}
2d(k-1)\frac{k-1}{d-1} - (k-2)(k-1)- d(d-1) \left (\frac{k-1}{d-1}\right)^2 
= \frac{D-2}{(k-1)(d-1)}.
\ee

\subsection{WKB analysis}

Next we do a WKB analysis of the
effect of short length variations of the conformal factor.  We 
insert the ansatz
\be
u \propto e^{dA/2\hbar}
\ee
into \eq{wfc}, and take the formal $\hbar\rightarrow 0$ limit, obtaining
\be \label{eq:WKB-eqn}
 \frac{d}{4} g^{ij} (\nabla_i A)\nabla_j(dA+2(k-2)\hbar B) = \frac{\hbar^2}{6}U + {\cal O}(\hbar^2),
\ee
where the extra term on the left hand side comes from the connection.  
Using the standard formulas, the curvature contribution to $U$ will go as
\be
U \sim \frac{(k-1)(k-2)}{2} e^{-2B} (\nabla B)^2 ,
\ee
so by taking $B\sim 1/\hbar$ we keep the growing term in the curvature, while
$C$ and the background value of $U$ can be neglected.
Thus we are effectively in the ``ground state''
regime of \S \ref{s:ground-state}.
This also makes
the $e^{-2B}$ factor scale in the same way as the metric $g^{ij} \sim e^{-2A}$.

The WKB estimate for this contribution to the effective potential is then
\be
V_{eff} \sim -2\int \sqrt{g} e^{dA} U 
\ee
While the integrand looks exponentially small, we should be careful as the
conformal factor can give us exponentially large $\sqrt{g}$.  
In a small region, we can approximate $A$ and $B$ by their linear parts
$A \sim a\cdot x$ and $B\sim b\cdot x$; then
\be
d a(da+2(k-2)b) = \frac{(k-1)(k-2)}{3} b^2 ,
\ee
while $\sqrt{g} \sim e^{kB}$.  We choose $b>0$, while out of the
two solutions for $a$, we choose the one with $a<0$ by the WKB ansatz.
One can then show that 
the overall exponent $da+(k-2)b$ is negative for all $k>2$.

While the computation looks similar to the linearized analysis, it is not the same
because it takes into account the exponentials in the integrand, and is thus nonlinear.

\subsection{Radially symmetric ansatz}

Another nonlinear test at short distances can be done by considering 
a radially symmetric conformal factor, say
\be
e^{2B} = (a^2+r^2)^\gamma ,
\ee
For $-2<\gamma<-1$, the
conformal factor becomes large at $r\sim a$, producing a region
of large volume and large positive curvature.  On the other hand, it becomes
small for $r>>a$, so one can patch such a region (or ``bubble'') into a general
manifold $X$.  This type of construction is used in the study of the Yamabe problem
\cite{yamabe}.

While we omit the details here, one can show that the effective potential \eq{Veff-final}
remains bounded below in these metrics as well.

\medskip
\noindent
To summarize this section, a variety of
simple modifications to the conformal factor
have the full nonlinear energy bounded below.  Since this would seem to be 
necessary for string/M theory compactification on $X$ to be well defined,
we conjecture that it is always true.  

From a physics point of view, it would be even more interesting if this worked for
some $X$ and failed for others, as this could give us a consistency condition and cut
down on the plethora of vacua.  The only evidence for this we see so far is that the
case $d=3$ might be special, since the source in \eq{C-constraint} has a negative
power of $u$ in this case.

While one might go on to conjecture that $V$ is bounded below for all metrics on
a given manifold $X$, there are some mathematical reasons to doubt this \cite{LeBrun}.
Physically it is not required, as long as $V$ has local minima, for which
the tunnelling rates to lower minima are very small (as in \cite{Kachru:2003aw}).

When $V$ is bounded below, the bound can be regarded as a topological invariant 
of $X$, directly analogous to the Yamabe invariant \cite{yamabe}.
Of course, it might be the same invariant, as turned out
to be the case for Perelman's entropy \cite{LeBrun}.

\section{Solutions}

So far, we avoided using the $k$-dimensional Einstein equation \eq{k-einstein-u}, 
so that we could derive an effective potential which is valid off-shell.  
In this section we generally assume we are expanding around a solution.

\subsection{Solving the $k$-dimensional equations}
\label{ss:k-dim}

We now
bring in the Einstein equation, reinterpreted as above:
\bea \label{eq:k-einstein-u-two}
u^2 \left(R^{(k)}_{ij} - \half g_{ij} R^{(k)} \right) - \nabla_i \nabla_j u^2
+ g_{ij} \nabla^2 u^2 &=& u^2 T^{(k)}_{ij} - \frac{4(d-1)}{d} \nabla_i u \nabla_j u \\
\nonumber
 &&- \half g_{ij} \left(  - \frac{4(d-1)}{d} (\nabla u)^2- C  u^{2-4/d} 
 \right) 
\eea
Note that $T^{(d)}_{string}$ does not appear in this formula, however since normally
it depends on the metric on $X$, terms from it will appear in $T^{(k)}_{ij}$.

Its trace is
\be \label{eq:k-trace-eqn}
\frac{k-2}{2} R^{(k)} = 2(k-1) u^{-1} \nabla^2 u + \frac{2(D-2)}{d}(u^{-1}\nabla u)^2
 -\frac{k}{2} u^{-4/d} C -T^{(k)} . 
\ee
Finally, the flux equations of motion are
\be
d*(u^2 F) = dF  = 0.
\ee

It might seem tempting to eliminate $C$ using the constraint \eq{d-trace}, to get an equation in terms
of $k$-dimensional quantities, equivalent to \eq{k-Einstein}.
However this is not a good idea, because we would lose the knowledge that $C$
is constant on $X$.

Rather, we can use \eq{d-trace} to eliminate either $R^{(k)}$ or $-\nabla^2 u$.
If we do the first, we get
\be \label{eq:nabla-u-equation}
 -(D-2)\nabla^2 u^2 +\left(dT^{(k)}-(k-2)T^{(d)} 
 \right) u^2=
(D-2) C u^{2-4/d}  .
\ee
Integrating this equation over the manifold gives
\be \label{eq:nabla-u-relation}
(D-2) C = - G_N \int \sqrt{g} \left(dT^{(k)}-(k-2)T^{(d)}
\right) u^2 ,
\ee
which is the constraint appearing in the no-go theorems of \cite{deWit,Maldacena-Nunez}.
Writing it in terms of $u^2$, 
it becomes a Schr\"odinger equation with a nonlinear source.

If we do the other elimination, we get
\be\label{eq:Rk-equation}
R^{(k)} = 
-\frac{4(d-1)}{d}(u^{-1}\nabla u)^2
 + u^{-4/d} C +\frac{2(d-1)}{D-2}T^{(k)} -\frac{2(k-1)}{D-2} T^{(d)} . 
\ee

Both are interesting equations.
The combinations of flux stress-tensors which appear are
\be
\alpha T^{(k)}   +\beta T^{(d)} = \half\sum_p ((2p-k)\alpha-d\beta)  |F^{(p)}|^2 .
\ee
In the original trace equation \eq{k-trace-eqn}, we have
\be
- T^{(k)}   -\frac{k}{2} T^{(d)} = \frac{1}{4}\sum_p (-4p+k(d+2))  |F^{(p)}|^2 ,
\ee
which for $k>4$ is always positive.
For \eq{nabla-u-equation}, we have
\be
dT^{(k)}-(k-2) T^{(d)} = d\sum_p (p-1)  |F^{(p)}|^2 .
\ee
As in the no-go theorems, this is non-negative except for $p=0$.
Finally, in \eq{Rk-equation} we have
\be
2(d-1)T^{(k)} -2(k-1) T^{(d)}=
 \sum_p ((d-1) 2p+k-d) |F^{(p)}|^2 
\ee
Again, for $k>d$ this is always positive.

To summarize, \eq{k-trace-eqn} and \eq{Rk-equation} make the point that flux favors positive scalar
curvature on $X$, while stringy corrections are needed to get negative scalar curvature.  We can now
go on to consider how the curvature is distributed in the internal dimensions, how this impacts
the warp factor, and whether the consistency condition $\lambda_1>0$ of the previous section
is significant or not.  We will look at this in examples in future work.

\iftrue
\subsection{Conformal factor}

Next, we introduce the conformal factor.  Fixing all the other fields, we would like to find
the conformal factor which minimizes $V_{eff}$.  Granting that this is bounded below, there
should be a minimum satisfying $\delta V_{eff}/\delta B=0$ (of course, there could be
other critical points).  We need to fix the volume modulus to have a minimum; rather than
do this physically we simply impose $\Vol(X)=\int \sqrt{g} v^{2k/(k-2)}$ with a Lagrange
multiplier $D$.

Taking $v=e^{(k-2)B/2}$, and $F=0$, we can rewrite \eq{Veff-final} as
\bea \label{eq:Veff-with-conformal}
V_{eff} &=&  \half\int \sqrt{g^{(k)}}  \bigg[ -u^2 v^2 R - \frac{4(k-1)}{k-2}(\nabla v)\nabla(u^2 v)
- \frac{4(d-1)}{d} v^2 (\nabla u)^2 \\
&&\qquad\qquad + \frac{u^2}{2} \sum_p v^{2(k-2p)/(k-2)} |F^{(p)}|^2 \bigg] \\
&&+ C  \left(\frac{1}{G_N} - \int \sqrt{g} v^{2+4/(k-2)} u^{2-4/d} \right)
+ D \left(\Vol(X) - \int \sqrt{g} v^{2+4/(k-2)} \right).
\eea
The various $p$-form flux energies come with different powers of
the conformal factor $v$, given by the $L$ scaling of \S \ref{s:physical},
while Chern-Simons terms do not depend on the conformal factor.
We will not quote these in the equations below, but they can be easily added.

We expect that $V_{eff}$ is minimized at the simultaneous critical point
\bea 
0 &=& \frac{\delta V_{eff}}{\delta u} = \nonumber
 -u v^2 R
 + \frac{4(k-1)}{k-2}\, uv\,\nabla^2 v
+ \frac{4(d-1)}{d} \nabla (v^2 \nabla u)
- \frac{d-2}{d} C  v^{2+4/(k-2)} u^{1-4/d} \\ \nonumber
0 &=& \frac{\delta V_{eff}}{\delta v} =
-u^2 v R  
+ \frac{2(k-1)}{k-2}\nabla^2(u^2 v)
+ \frac{2(k-1)}{k-2}\nabla(u^2 \nabla v) \\ \nonumber
&&\qquad \qquad - \frac{4(d-1)}{d} v (\nabla u)^2 
- \frac{k}{k-2} v^{1+4/(k-2)} \left( C u^{2-4/d} + D \right).
\eea

\fi

We can simplify the $u$ constraint a bit by the change of variables $u=w/v$, to get
\bea \label{eq:w-constraint}
0 &=& 
 - w v R
 - \left(\frac{4(d-1)}{d} -\frac{4(k-1)}{k-2}\right)\, w\,\nabla^2 v
+ \frac{4(d-1)}{d} v \nabla^2 w 
- \frac{d-2}{d} C  v^{2+4/(k-2)} (w/v)^{1-4/d} \\
0 &=& \nonumber
-w^2  R  
+ \frac{2(k-1)}{k-2}\, v\nabla v^{-1}\nabla(w^2)
- \frac{4(d-1)}{d}  (\nabla w - w\nabla\log v)^2 
- \frac{k}{k-2} v^{2+4/(k-2)} \left( C (w/v)^{2-4/d} + D \right). \nonumber
\eea

Except for the source term, the constraint is now linear in $v$ and $w$ separately.

\subsection{Comparison with supersymmetric ansatzes}

Any supersymmetric ansatz should be using the conformal factor which
minimizes the potential, so let us sketch how this works in two well-studied examples.

In \cite{Becker:1996gj}, K. and M. Becker constructed M theory compactifications
on Calabi-Yau fourfolds, thus $d=3$ and $k=8$.  They take
$B=-A/2$ (see their Eqs. (2.11) and (2.30)), so $v=u^{-1}$.  Their construction involves
both electric and magnetic four-form flux, so in the magnetic notation uses $p=4$ and $p=7$.
Stringy ingredients include the Chern-Simons term, and an $R^4$ anomaly term.

In \cite{Giddings:2001yu}, Giddings, Kachru and Polchinski developed
supersymmetric type IIb compactification on Calabi-Yau threefolds
with flux.  Here $d=4$ and  $k=6$.  We will consider the special case in which the dilaton is constant.
They take $B=-A$, so again $v=u^{-1}$.  This construction uses $p=3$ and $p=5$ flux, a
Chern-Simons term $F\wedge H\wedge C^{(4)}$, and orientifold sources to $T^{(d)}_{string}$.

The two constructions are similar enough that we can make a unified discussion.
Both are based on Calabi-Yau manifolds, so $R=0$.  Both lead to Minkowski space-time,
so we can set $C=0$.  We note in passing that the ground state approximation of 
\S \ref{s:ground-state} is always exact in this case.

Both constructions
use conformally invariant flux with $2p=k$ (call this $F_a$), 
and an additional $p>k/2$ magnetic flux (call this $F_b$) which
is determined by the warp factor,
$F_b= *^{(D)}\epsilon^{(d)}\wedge d(u^2)$, where $*^{(D)}$ is the $D$-dimensional Hodge star
in the warped metric.\footnote{This follows from the cancellation between warped tension and
potential energy for supersymmetric space-time filling M2 or D3 branes.}
Substituting the ansatz, in both this becomes $F_b= *^{(k)} d(v^2)$.

In both cases, the function $w$ of \eq{w-constraint} is set to a constant (say $1$), so
this equation becomes linear.  For Minkowski space-time, $C=0$ and one has an integral
constraint on $T^{(d)}$, which in some sense is a supersymmetric partner to a 
topological constraint on $F_a^2-T^{(d)}_{string}$
(the M2 or D3 tadpole condition).

Let us derive the equations of motion by re-expressing the
effective potential \eq{Veff-with-conformal} in terms of $v$ and $w$,
\be
V_{eff} =  \half\int \sqrt{g^{(k)}}  \bigg[  -\frac{4(k-1)}{k-2}(\nabla v)\nabla(w^2/ v)
- \frac{4(d-1)}{d} v^2 (\nabla (w/v))^2 
 + \frac{w^2}{2v^2} (F_a^2 + v^{-2} F_b^2) \bigg]
\ee
There is also a term $\frac{w^2}{2v^2} T^{(d)}_{string}$, but since this always comes
with $F_a^2$ and has the same dependence on $v$ and $w$, we leave it out until the end.

After varying $V_{eff}$, we set $w=1$, to find
\bea \label{eq:foo}
0 &=& \frac{\delta V_{eff}}{\delta w} =
\alpha\frac{1}{v}\nabla^2 v
 + \frac{1}{2v^2} (F_a^2 + v^{-2} F_b^2) \\
0 &=& \frac{\delta V_{eff}}{\delta v} =
-\alpha
(v^{-1} \nabla^2 v - v^{-2} (\nabla v)^2)
 - \frac{1}{2v^2} (F_a^2 + 2v^{-2} F_b^2)
\eea
where 
$$\alpha\equiv\left(\frac{4(k-1)}{k-2}-\frac{4(d-1)}{d}\right) > 0.$$

Subtracting these equations, we find
\be
\alpha(\nabla  v)^2 =
 \frac{1}{2v^2}  F_b^2 ,
\ee
which is satisfied by $F_b=\sqrt{\alpha/2}\nabla v^2$.  Substituting in \eq{foo}, 
and restoring $T^{(d)}_{string}$, we find
$$
-\alpha\nabla^2 v^2 =   F_a^2 - T^{(d)}_{string},
$$
which is the warp factor equation in both cases.

Two points worth observing are the supersymmetry relation between the warp factor and $F_b$,
and how the positive curvature required by
\eq{k-trace-eqn} is provided by the variation of the conformal factor $v$.  

It would be very interesting to find a derivation of the effective superpotential along
the lines of \S \ref{s:general},
starting with the $D$-dimensional supergravity Lagrangian and an unbroken
supersymmetry, and producing a functional of the $k$-dimensional fields.

\vskip 0.2in
{\it Acknowledgements}

This work grew out of ongoing discussions with Renata Kallosh on compactification
using negatively curved manifolds, to whom we give special thanks.  Also influential
were many discussions over the years with Tom Banks on problems with the effective potential in
gravity and string theory.
We also thank Michael Anderson,
Xiuxiong Chen, Gary Gibbons, Marcus Khuri,
Luca Mazzucato, John Morgan, Joe Polchinski, Dennis Sullivan,
Peter van Nieuwenhuizen and Edward Witten
for valuable discussions.  Finally, we thank Steve Giddings for pointing out 
\cite{Giddings:2005ff} and subsequent discussions.

This research was supported in part by DOE grant DE-FG02-92ER40697.

\appendix

\section{Conventions}

Supergravity considerations will be accurate
when all geometric length scales, such
as the diameter and volume of $X$ and curvature lengths, are greater
than $1$ in fundamental units (we ignore the string coupling), so we take
$M_{Planck,D}=1$ to make this condition evident.  We furthermore take
the coordinates on $X$ to range over order $1$, so a geometric length
scale $L$ goes as $L^2 \sim  g_{ij}$.
Then $\Vol X \sim \sqrt{g} \sim L^k$, derivatives $\nabla \sim 1$, and the
scalar curvature $R \sim 1/L^2$.
The integrals of the $p$-form fluxes over homology cycles $\Sigma^{(p)}$
are quantized as 
\be
N_p \sim \int_{\Sigma^{(p)}} F^{(p)} \sim F \Vol \Sigma^{(p)}
\ee
Thus $\int \sqrt{g}|F^{(p)}|^2 \sim  N_p^2 L^{k-2p}$ in fundamental units, and $T_{string} \sim 1$.

\section{Computations of curvature}

Following Wald appendix D, we write $\nabla_A$ for
the covariant derivative adapted to the metric $g$, and $\tnabla_A$ for that
adapted to $\tg$, so
\be
C^c_{~ab} \equiv \nabla_a \omega_b - \tnabla_a \omega_b .
\ee
satisfies
\be
C^c_{~ab} = \half \tg^{cd}\left( \nabla_a \tg_{bd} +\nabla_b \tg_{ad} -\nabla_d \tg_{ab} \right).
\ee
The curvature of $\tg$ is then (Wald 7.5.8)
\be
\tR_{abc}^{~~~d} = R_{abc}^{~~~d} - 2 \nabla_{[a} C^d_{~b]c} + 2 C^e_{~c[a} C^d_{~b]e}.
\ee
Note that \eq{metric} has
a $\IZ_2$ symmetry under $e^\mu\rightarrow -e^\nu$, which
constrains the connection and curvature coefficients.
Using $\nabla_a g_{bc}=R_{ij\mu}^{~~~\nu}=0$, we find
\begin{eqnarray}
C^i_{~jk}           &=& \delta^i_j \partial_k B + \delta^i_k \partial_j B - g_{jk} \nabla^i B \\
C^\mu_{~\nu i} &=& \delta^\mu_\nu \partial_i A \\
C^i_{~\mu\nu}  &=& -e^{2A-2B} \eta_{\mu\nu} \nabla^i A \\
\tR_{ijk}^{~~~l}     &=& R_{ijk}^{~~~l} 
 + 2\delta^l_{[i} \nabla_{j]} \nabla_k B
 - 2 g_{k[i} \nabla_{j]} \nabla^l B \\
&&  + 2 (\nabla_{[i} B) \delta^l_{j]} \nabla_k B
  - 2 (\nabla_{[i} B) g_{j]k} \nabla^l B
  - 2 g_{k[i} \delta^l_{j]} |\nabla B|^2 \nonumber
 \\
\tR_{ij\mu}^{~~~\nu} &=& - 2 \nabla_{[i} C^\nu_{~j]\mu} + 2 C^\lambda_{~\mu [i} C^\nu_{~j]\lambda} \\
&=& 0 \qquad (\sim \nabla_{[i}\nabla_{j]} A + \nabla_{[i} A \nabla_{j]} A) \nonumber
  \\
\tR_{i\mu j}^{~~~\nu} &=& -  \nabla_{i} C^\nu_{~\mu j} +  C^k_{~j i} C^\nu_{~\mu k}  
 - C^\lambda_{~j \mu} C^\nu_{~i \lambda} \nonumber\\
 &=&  \delta^\nu_{\mu} \left(-\nabla_i \nabla_j A 
  + C^k_{ij} \nabla_k A - (\nabla_i A)(\nabla_j A)  \right) \label{eq:imujnu}\\
 &=&  \delta^\nu_{\mu} \left(-\nabla_i \nabla_j A 
  - \nabla_i (A-B) \nabla_j (A-B)  + \nabla_i B\nabla_j B
   - g_{ij}(\nabla A\cdot\nabla B)  \right) \nonumber\\
\tR_{\mu\nu\lambda}^{~~~~\rho} &=& R_{\mu\nu\lambda}^{~~~~\rho}
 + 2 C^i_{~\lambda[\mu} C^\rho_{~\nu]i} \\
 &=& R_{\mu\nu\lambda}^{~~~~\rho}
 - 2 e^{2A-2B} \eta_{\lambda[\mu}  \delta^{~\rho}_{\nu]} (\nabla A)^2 \nonumber
\end{eqnarray} 
The other mixed components can be obtained using the antisymmetry $R_{[ab][cd]}$.
Here $\tR_{ijk}^{~~~l}$ is the same as for a $k$-dimensional conformal transformation
(Wald D.7).

One check is the case $A=B$.  The only nontrivial one is
\be
\tR_{i\mu j}^{~~~\nu}     =
 \delta^\nu_{\mu} \left(-\nabla_{i} \nabla_j B
  +  \nabla_{i} B \nabla_j B
  -  g_{ij} |\nabla B|^2 \right)
\ee
from the standard formula, which agrees with \eq{imujnu} at $A=B$.

Contracting indices, we find
\bea
\tR_{ik} = R_{ik} &
 -(k-2)\nabla_i\nabla_k B
 - g_{ik} \nabla^2 B
  + (k-2)\nabla_{i} B \nabla_k B
  -(k-2) g_{ik} (\nabla B)^2 \\
   & + d \left(-\nabla_i \nabla_k A 
  - \nabla_i (A-B) \nabla_k (A-B)  + \nabla_i B\nabla_k B \nonumber
   - g_{ik}(\nabla A\cdot\nabla B)  \right) \\
\tR_{\mu\lambda} = R_{\mu\lambda} &
   + e^{2A-2B} \eta_{\mu\lambda} \left( - d (\nabla A)^2
    - \nabla^2 A - (k-2) (\nabla A\cdot \nabla B)
   \right)
\eea
Again, this agrees with the standard formula for $A=B$.

Finally, the scalar curvature is a sum of two terms obtained by tracing these two contributions.
Let us define the Laplacian with respect to the conformally transformed 6d metric,
\bea
\tDelta &&\equiv -\frac{1}{\sqrt{\tg}}\partial_i \sqrt{\tg} \tg^{ij} \partial_j \\
 &=& -\nabla^2 -(k-2)\nabla B\cdot\nabla ,
\eea
then
\bea
\tR^{(k)} &=& e^{-2B} \left( R^{(k)}  \label{eq:scalarRk}
 -(2k-2)\nabla^2 B
  -(k-2) (k-1) (\nabla B)^2 
    + d (\tDelta A - (\nabla A)^2  ) \right)\\
\tR^{(d)} &=& e^{-2A} R^{(d)}
   + d e^{-2B} \left( \tDelta A  - d (\nabla A)^2  \right)
\eea

Note that the term $\tDelta A  - d (\nabla A)^2$ is precisely the scalar Laplacian associated
to the metric \eq{metric}, restricted to functions independent of space-time.  Thus, it is
a total derivative, and its contribution to $\int \sqrt{\tg} \tR^{(d)}$ vanishes (of course
the other term still depends on $A$).  

Finally, the conformally invariant Laplacian is $\Delta_c=\Delta + \alpha R$ with
\be
s = -\frac{k-2}{2}; \qquad \alpha = \frac{k-2}{4(k-1)};
\ee
such that
\be \label{eq:conf-Lap}
\tDelta_c e^{sB} \phi = e^{(s-2)B} \Delta_c  \phi.
\ee


\end{document}